\documentclass[letterpaper, 10 pt, conference]{templates_itsc/ieeeconf}  % Comment this line out if you need a4paper

\IEEEoverridecommandlockouts                              % This command is only needed if 
\usepackage{amsmath,amssymb,amsfonts}
\usepackage{algorithmic}
\usepackage{graphicx}
\usepackage{acro}
\usepackage{textcomp}
\usepackage{xcolor}
\acsetup{patch/maketitle=false}
\DeclareAcronym{RL}{
    short = RL,
    long = reinforcement learning
}

\DeclareAcronym{SAC}{
    short = SAC,
    long = Soft Actor-Critic
}

\DeclareAcronym{HER}{
    short = HER,
    long = Hindsight Experience Replay
}

\DeclareAcronym{HRL}{
    short = HRL,
    long = hierarchical reinforcement learning 
}

\DeclareAcronym{MDP}{
    short = MDP,
    long = Markov decision process 
}

\DeclareAcronym{ADS}{
    short = ADS, 
    long = Automated Driving System
}

\DeclareAcronym{DSL}{
    short = DSL, 
    long = Domain Specific Language
}

\DeclareAcronym{ADAS}{
    short = ADAS, 
    long = Advanced Driver Assistance System
}

\DeclareAcronym{ALKS}{
    short = ALKS, 
    long = Active Lane Keeping Systems
}

\DeclareAcronym{SUT}{
    short = SUT, 
    long = System under Test
}

\DeclareAcronym{LSTM}{
    short = LSTM,
    long = long short-term memory
}

\DeclareAcronym{NURBS}{
    short = NURBS,
    long = Non-Uniform Rational B-Splines
}

\DeclareAcronym{GOOSE}{
    short = GOOSE,
    long = Goal-conditioned Scenario Generation
}

\usepackage{bm}
\usepackage{algorithmic}
\usepackage{algorithm}
\usepackage{siunitx}

\title{\LARGE \bf
GOOSE: Goal-Conditioned Reinforcement Learning for Safety-Critical Scenario Generation
}

\author{Joshua Ransiek$^{1}$, Johannes Plaum$^{2}$, Jacob Langner$^{1}$ and Eric Sax$^{3}$% <-this % stops a space
\thanks{$^{1}$Joshua Ransiek and Jacob Langner are with the FZI Research Center for Information Technology, Karlsruhe, Germany
        {\tt\small ransiek@fzi.de}}%
\thanks{$^{2}$Johannes Plaum is with Torc Europe GmbH, Stuttgart, Germany
        {\tt\small johannes.plaum@torc.ai}}%
\thanks{$^{2}$Eric Sax is with the KIT Karlsruhe Institute of Technology, Karlsruhe, Germany
        {\tt\small eric.sax@kit.edu}}%
}

\begin{document}

\maketitle
\thispagestyle{empty}
\pagestyle{empty}

\begin{abstract}
Scenario-based testing is considered state-of-the-art for verifying and validating \acp{ADAS} and \acp{ADS}.
However, the practical application of scenario-based testing requires an efficient method to generate or collect the scenarios that are needed for the safety assessment. In this paper, we propose \ac{GOOSE}, a goal-conditioned \ac{RL} approach that automatically generates safety-critical scenarios to challenge \acp{ADAS} or \acp{ADS} (Fig. \ref{fig:overview}). In order to simultaneously set up and optimize scenarios, we propose to control vehicle trajectories at the scenario level. Each step in the \ac{RL} framework corresponds to a scenario simulation. We use \ac{NURBS} for trajectory modeling. To guide the goal-conditioned agent, we formulate test-specific, constraint-based goals inspired by the OpenScenario \ac{DSL}. Through experiments conducted on multiple pre-crash scenarios derived from UN Regulation No. 157 for \ac{ALKS}, we demonstrate the effectiveness of GOOSE in generating scenarios that lead to safety-critical events.
\end{abstract}    
\section{Introduction}
\label{introduction}

Scenario generation is one of the essential steps in scenario-based testing and, therefore an important part of the verification and validation of \acp{ADAS} and \acp{ADS}. The core principle in scenario-based testing is to use scenarios with a predefined starting parameterization which are then simulated using the \ac{SUT} in a closed-loop simulation \cite{riedmaier2020survey}. 
This approach differs from distance-based or statistical testing, which relies on the vehicle driving on the road to encounter a sufficient number of critical and non-critical situations to perform safety assessments.
According to \cite{schutt20231001}, most scenario acquisition methods require predefined base scenarios that can be altered.
This implies, dangerous and critical situations have to be known before the testing phase. Nevertheless, not all critical scenarios needed for safety assessment may be known initially, which is why the ISO 21448 standard explicitly demands to minimize the risk of unknown critical scenarios \cite{InternationalOrganizationforStandardization2022}. Scenario generation without the boundaries of a base scenario tries to address this problem.

Current practices for scenario generation in the scenario-based testing predominantly rely on engineers or involve a semi-automated process. In the semi-automated approach, a human operator specifies the number of actors, their initial positions, and predetermined trajectories, while their speeds are selected from a range of options. Human involvement can be time-consuming and may lead to critical scenarios being overlooked. To address this challenge, researchers are actively exploring the use of agents in simulations to create collisions or exhibit adversarial behavior. However, these agents often require training to maneuver vehicles effectively, and the generated scenarios may surpass the system's capabilities, resulting in unrealistic situations.

\begin{figure}[t!]
    \centering
    \vspace{2mm}
    \includegraphics[width=0.99\linewidth]{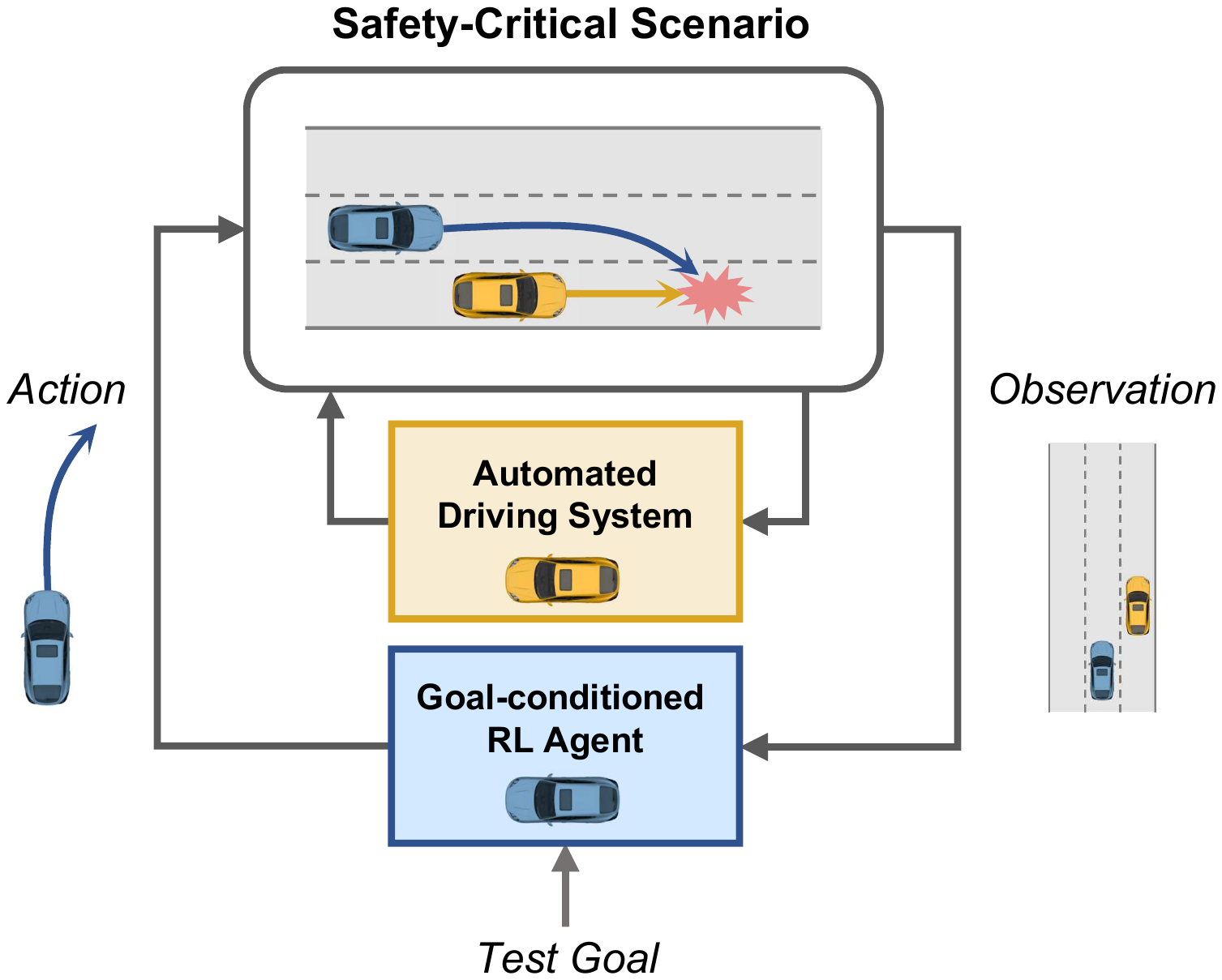}
    \vspace{-6mm}
    \caption{The Pipeline for safety-critical scenario generation using a goal-conditioned RL agent. The agent (blue) attempts to increase the criticality of the simulated scenario by modifying the adversarial vehicle's trajectory, while the ego vehicle (yellow) is trying to maintain a safe state.}
    \label{fig:overview}
\end{figure}
In this paper, we present a comprehensive approach to scenario generation, \textbf{Go}al-c\textbf{o}nditioned \textbf{S}cenario G\textbf{e}neration (GOOSE), for testing of \ac{ADAS} and \ac{ADS}. Our contributions consist of the following: 
1) an application of a goal-conditioned \ac{RL} approach using a constrained-based goal selection strategy for scenario generation, 
2) the utilization of \ac{NURBS} for efficient trajectory modeling and action space reduction of the goal conditioned policy, and 
3) experimental results on a selection of pre-crash scenarios show that \ac{GOOSE} is capable of generating safety-critical scenarios. 
\section{Related Work}

\subsection{Reinforcement Learning for Software Testing} 
Designing automated tests using \ac{RL} agents heavily focus on exploring new states by interacting with the software. \cite{nie2021play} employ \ac{RL} agents to identify bugs in games. Similarly, \cite{zheng2019wuji} use a state-counting method to encourage exploration and \cite{gordillo2021improving,zhan2018taking} use a curiosity objective. In \cite{borovikov2019winning} \ac{RL} is applied in a game production setting for game evaluation and balancing purposes. Furthermore, agents are utilised to explore graphical user interfaces (GUIs) of Android applications \cite{eskonen2020automating, koroglu2018qbe, pan2020reinforcement} and computer software \cite{harries2020drift}, while \cite{ahmad2019exploratory, koo2019pyse} employ agents to identify performance bottlenecks in server applications.  

\subsection{Goal-conditioned Reinforcement Learning} 
Goal-conditioned \ac{RL} \cite{andrychowicz2017hindsight, kaelbling1993learning, nachum2018data, schaul2015universal} aims to train agents that can reach
any goal provided to them. Given the current state and the goal, the resulting goal-conditioned policy predicts action sequences that lead towards the desired goal. \Ac{HER} \cite{andrychowicz2017hindsight, kaelbling1993learning} is often used to enhance the robustness and sample efficiency of goal-conditioned policies by relabeling or reweighting the reward obtained from an achieved goal. \Ac{HRL} addresses the goal-conditioned setting by learning a higher-level policy to predict a sequence of intermediate subgoals iteratively. These subgoals can then serve as targets for low-level policies \cite{chane2021goal, nachum2018data,nasiriany2019planning}. 

\subsection{Safety-Critical Traffic Scenario Generation} 
The development of \acp{ADS} necessitates the creation of diverse and, ideally, realistic scenarios to ensure safe and efficient operation of the system in various traffic situations. 
The majority of approaches for safety-critical scenario generation involve perturbing the maneuvers of interactive actors in an existing scenario with adversarial behaviors.
\cite{rempe2022generating} modifies real world scenes in the latent space of a learned traffic model based on a graph-based conditional variational autoencoder (VAE). 
Scenario optimization through differentiable simulation by learning from demonstration and common sense is considered in \cite{suo2021trafficsim}. 
\cite{tan2021scenegen} propose a neural autoregressive model to insert actors of different classes into traffic scenes. 
\cite{wang2021advsim} modify the trajectories of the actors in a physically plausible way using query-based black-box optimisation. 
Furthermore, \cite{wiederer2022benchmark} use a human controller to create abnormal scenarios in the simulation. 
Prior works already choose \ac{RL} to solve the safety-critical scenario generation task. 
\cite{koren2018adaptive,lee2020adaptive} introduce adaptive stress testing (AST) for automotive applications in a crosswalk scenario. 
AST is extended using reward augmentation \cite{corso2019adaptive}, a \ac{LSTM} policy network \cite{koren2019efficient}, the GoExplore algorithm \cite{koren2020adaptive}, and a combination of backward algorithm and proximal policy optimization (PPO) \cite{koren2021finding}. 
Similarly, \cite{karunakaran2020efficient} employ a \ac{RL} agent in a pedestrian crossing to evaluate a collision avoidance system. 
Various approaches \cite{kuutti2020training,chen2021adversarial,sun2021corner} use adversarial agents in highway scenarios. \cite{o2018scalable} use a risk-based framework and \cite{feng2021intelligent} train vehicles to execute adversarial maneuvers to increase the collision probability. \cite{niu2023re} learn adversarial policies from both offline naturalistic driving data and online simulation samples. 
\cite{haq2023many} incrementally generate sequences of environmental changes by combining \ac{RL} and many-objective search in straight, left-turn, and right-turn scenarios. In \cite{ding2020learning} the safety of a driving algorithms is evaluated in an intersection scenario. A method for finding failure scenarios that trains the adversarial agents using multi-agent RL is proposed by \cite{wachi2019failure}. 

In this work the goal-conditioned \ac{RL} approach is applied to a scenario generation process where the goals define critical events and desired scenario properties. The idea of identifying software bugs by exploring new states is employed to generate safety-critical scenarios for a given \ac{ADAS} or \ac{ADS}.
In contrast to existing approaches, we will not train interactive agents to create collisions or exhibit adversarial behaviour. 
This is due to the fact that these agents require training in order to operate vehicles in an effective manner. Furthermore, their behaviour may exceed the system’s capabilities, resulting in unrealistic situations.
Adopted from previous work \cite{ransiek2023generation}, \ac{GOOSE} controls the vehicle trajectory at the scenario level, without the need for the vehicle to learn individual control. 
Each step in the \ac{RL} framework is equivalent to a
scenario simulation, which allows for the effective generation of challenging scenarios.

\section{Background}

\subsection{Goal-conditioned Markov Decision Process}

The goal-conditioned \ac{RL} problem is defined by a finite-horizon, goal-conditioned \ac{MDP} $(\mathcal{S},\mathcal{G},\mathcal{A},p,R,T_{max},\gamma, \rho_0,\rho_g)$, where $\mathcal{S}$ is the set of states, $\mathcal{G}$ is the set of goals, $\mathcal{A}$ is the set of actions, $p(\bm{s}_{t+1}\,|\,\bm{s}_t,\bm{a}_t)$ is the time-invariant dynamics function, $R$ is the reward function, $T_{max}$ is the maximum horizon, $\gamma$ is the discount factor, $\rho_0$ is the initial state distribution, and $\rho_g$ is the goal distribution \cite{schaul2015universal}. 
The objective in goal-conditioned \ac{RL} is to obtain a policy $\pi(\bm{a}_t\,|\,\bm{s}_t,\bm{g})$ to maximize the expected sum of rewards 
\begin{equation}
    J(\pi) = \mathbb{E}\Bigg[\sum^{T_{max}}_{t=0} \gamma^t R(\bm{s}_t,\bm{a}_t,\bm{g})\Bigg], 
\end{equation}
where the goal is sampled from $\rho_g$ and the states are sampled according the $\mathbf{s_0} \sim \rho_0$, $\bm{a}_t \sim \pi(\bm{a}_t \,|\, \bm{s}_t, \bm{g})$, and $\bm{s}_{t+1} \sim p(\bm{s}_{t+1}\,|\,\bm{s}_t,\bm{a}_t)$.  

\subsection{Non-Uniform Rational B-Splines}
\begin{figure}[!b]
    \centering
    \includegraphics[width=0.99\linewidth]{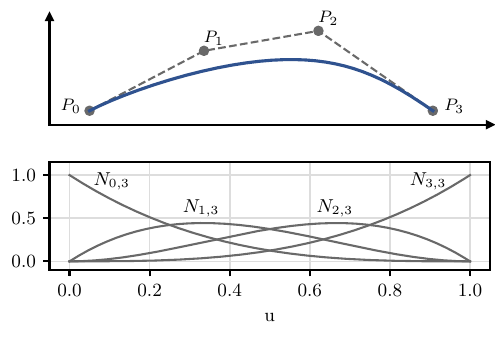}
    \caption{Exemplary NURBS curve of 3-th degree with weights $\omega_{0,...,3} = 1$ and the associated basis functions $N_{\{0,...,3\},3}$}
    \label{fig:nurbs_example}
\end{figure}
\label{subsec:NURBS}
The \ac{NURBS} curve represents a parametric curve $C(u)$ that uses the B-spline functions as its basis \cite{piegl2012nurbs}. 
The formulation of the curve is shown in Equation (\ref{eq:bezier_curve}).
\begin{equation} \label{eq:bezier_curve}
    C(u) = \frac{\sum^{n}_{i=0} N_{i,p}(u) \omega_i P_i}{\sum^{n}_{i=0} N_{i,p}(u) \omega_i}, \;\;\; a\leq u \leq b,
\end{equation}
where, $p$ represents the curve degree, $n$ represents the number of control points, $P_i$ represents the $i$-th control point, $\omega_i$ represents the weight of the $i$-th control point, and $N_{i,p}(u)$ represents the $p$-th degree B-spline basis function defined on the non-periodic knot vector 
\begin{equation}
    U = \{\underbrace{a,...a}_{p+1},u_{p+1},..., u_{m-p-1},\underbrace{b,...,b}_{p+1}\}.
\end{equation}
The number of knots $m + 1$ is calculated by $m = p + n + 1$. The $i$-th B-spline basis function of $p$-th degree is defined as 
\begin{equation}
    N_{i,0}(u) = 
    \begin{cases} 
    1 & \textrm{if} \,\, u_i \leq u < u_{i+1} \\
    0 & \textrm{otherwise}
    \end{cases},
\end{equation}
\begin{multline}
    N_{i,p}(u) = \frac{u-u_i}{u_{i+p}-u_i}N_{i,p-1}(u)\\ +\frac{u_{i+p+1}-u}{u_{i+p+1}-u_{i+1}}N_{i+1,p-1}(u),
\end{multline}
where $u_i$ is the $i$-th knot, and the half-open interval $[u_i,u_{i+1})$ is the $i$-th knot span. We use a cubic \ac{NURBS} curve (example shown in Figure \ref{fig:nurbs_example}), which is empirically found sufficient for modeling diverse trajectories, velocities and acceleration processes, while retaining a relatively small set of optimizable parameters. 
\section{Method}

\begin{figure}[!t]
    \vspace{1mm}
    \centering
    \includegraphics[width=0.90\linewidth]{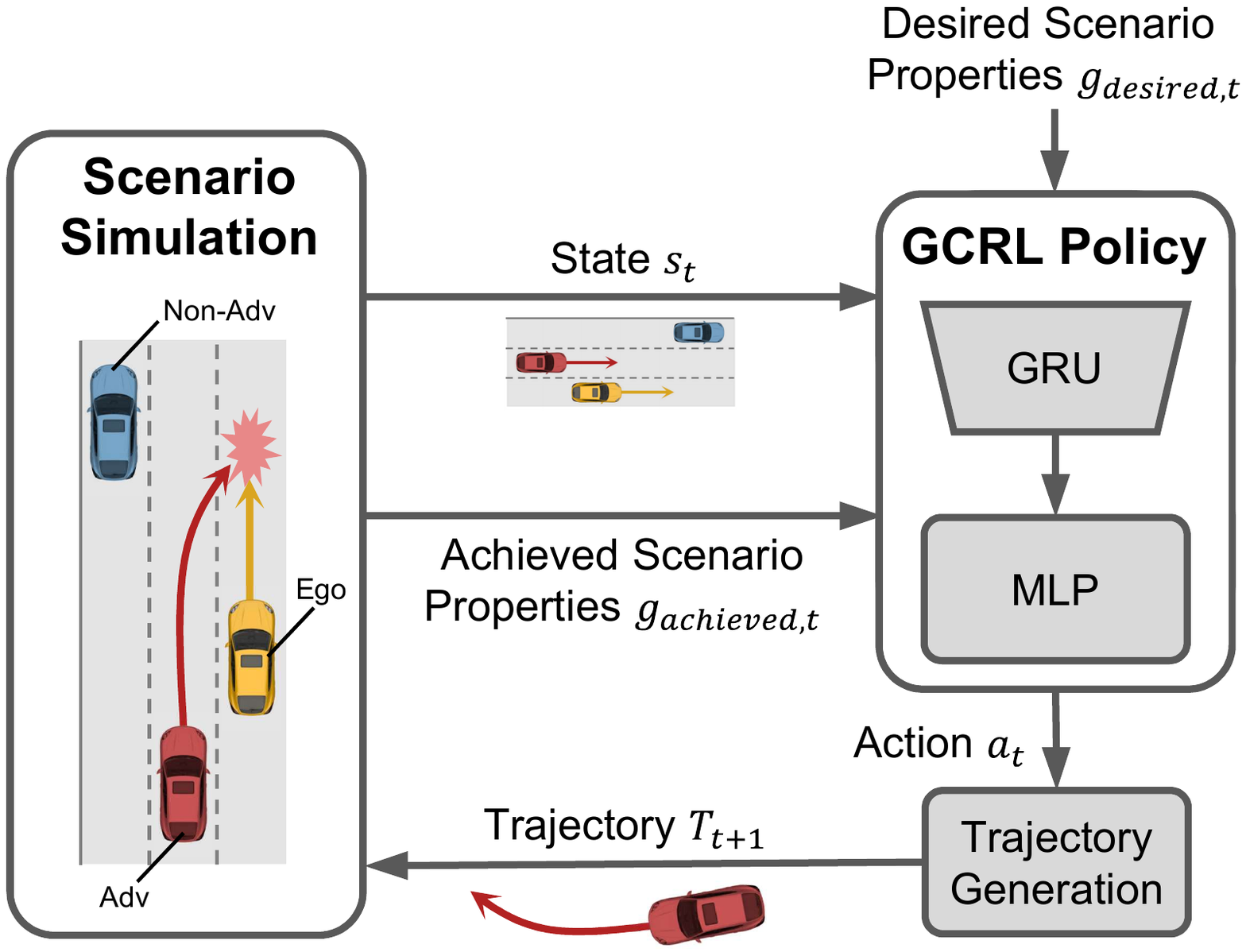}
    \caption{Method for scenario generation using a goal-conditioned RL agent. The agent (red) attempts to increase the criticality of the simulated scenario by modifying the target vehicle's trajectory based on state and goal, while the ego vehicle (yellow) is trying to maintain a safe state.}
    \label{fig:structure}
\end{figure}

The objective is to provide a \ac{RL} agent that exhibits adversarial behavior in the given scenario to generate safety-critical scenarios. We approach this problem by formulating it as an optimization problem in the framework of goal-conditioned \ac{RL}. The interaction between the \ac{RL} agent and the environment, which includes the scenario simulation and \ac{SUT}, is illustrated in Fig. \ref{fig:structure}. When the agent interacts with the environment at time step $t$, it receives state $\bm{s}_t$ that provides relevant information about the movement of the actors in the scenario. In addition, the agent is provided with a desired goal $\bm{g}_{desired,t}$, which specifies the desired scenario properties, and an achieved goal $\bm{g}_{achieved,t}$, which characterizes the most recently generated scenario. The distance between the achieved and the desired goal serves as an assessment metric for the generated scenarios and functions as a guidance for training the goal-conditioned policy. The trajectory of the target vehicle is incrementally altered by the agent's actions $\bm{a}_t$. Afterwards the environment proceeds to the next state $\bm{s}_{t+1}$, through the simulation of the generated scenario. The new scenario configuration results in a new ego vehicle trajectory, which is subsequently observed by the agent for further decision-making. We are particularly interested in scenarios where the \ac{SUT}, that controls the ego vehicle, must intervene and deviate from its comfortable trajectory. The system is considered a black-box, and the \ac{RL} agent has no access to its internal workings.

\ac{GOOSE} consists of three components, which are discussed in more detail in the following sections. These include the action space definition using NURBS (Sec. \ref{subsec:algo_action_space}), which is an efficient method for modeling entire vehicle trajectories, feature extraction from high-dimensional observations (Sec. \ref{subsec:algo_observation_space}) to preprocess the high-dimensional scenario observations, and the constraint-based goal definition (Sec. \ref{subsec:algo_goal_space}) to specify the scenario generation task. The algorithm is summarized in Sec. \ref{subsec:algo_summary}.

\subsection{Action Space Definition}
\label{subsec:algo_action_space}

\begin{figure}[!t]
    \vspace{1mm}
    \centering
    \includegraphics[width=0.99\linewidth]{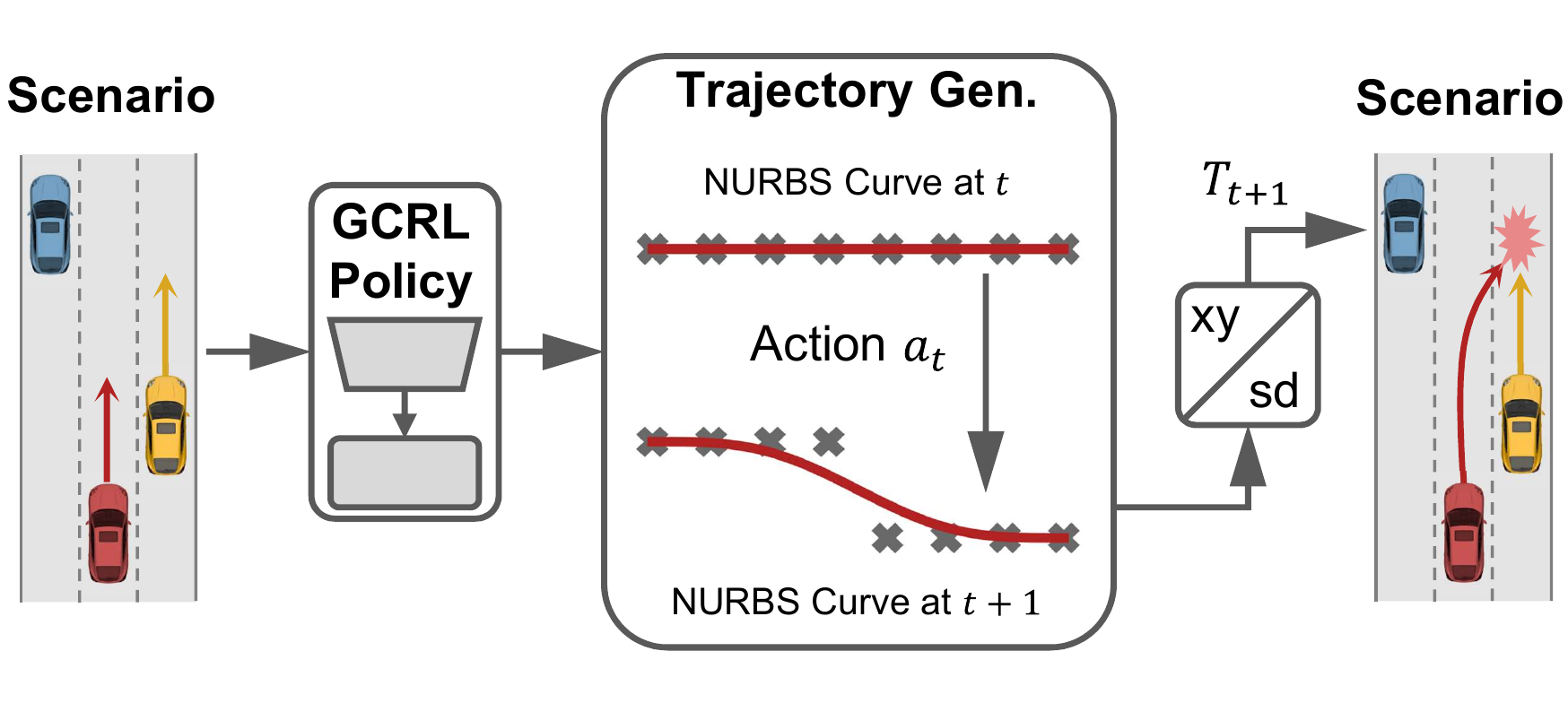}
    \caption{Scheme outlining the application of an action to a vehicle trajectory}
    \label{fig:action_space}
    \vspace{-3mm}
\end{figure}

The agent's actions permit it to alter the state of the environment, thereby modifying the scenario configuration.
To make the decision making invariant to the road curvatures, \ac{GOOSE} operates in the Frenet frame \cite{stoker1969differential}. 
On the Frenet space, the adversarial vehicle’s trajectory is specified in terms of longitudinal displacement $s$ and the lateral offset $d$. 
To further simplify the action space the high-dimensional continuous space of trajectory values is represented as \ac{NURBS} curve. 
The agent acts with the environment by incrementally altering the control points $P_i$ and weights $w_i$ of a \ac{NURBS} curve (Sec. \ref{subsec:NURBS}). 
Thus, the set of actions consists of three components. The incremental alteration of the longitudinal displacement $\bm{a}_{s}$, lateral offset $\bm{a}_{d}$ and weighting $\bm{a}_{\omega}$ of each control point $P$: 
\begin{equation}
\bm{a} = 
    \begin{bmatrix}
        \bm{a}_{s}& 
        \bm{a}_{d}& 
        \bm{a}_{\omega}
    \end{bmatrix}
\end{equation}
The $s$ and $d$ values, required for trajectory calculation in the Frenet frame, can be retrieved in a favorable discretization through point-wise evaluation of the \ac{NURBS} curve. 
The trajectory $T^{adv}_{sd}$ is then transformed into the Cartesian coordinates $T^{adv}_{xy}$ and used to update the movement of the adversarial vehicle. 
The new scenario configuration is provided to the \ac{SUT} to calculate a new ego vehicle trajectory $T^{ego}_{xy}$, which is subsequently observed by the deep \ac{RL} agent for further decision-making. The application of these actions is illustrated in Fig. \ref{fig:action_space}.

\subsection{State Space Definition}
\label{subsec:algo_observation_space}

The preprocessed state representation $\bm{z}$ is a tuple $(f,g_{a},g_{d})$, where $f$ is a feature vector of the observed scenario, $g_{a}$ is the achieved goal in the current step, and $g_{d}$ is the desired goal in the current episode. The set of possible states $S$ is vast and encompasses all relevant information about the movement of the actors within the scenario, and is expanded with the different goal definitions. Therefore, generalization is important to design an effective state representation.
The feature vector $f$ is extracted from the current scenario using a gated recurrent unit (GRU). The current scenario is represented by the following scenario observation:
\begin{equation}
\bm{s} = 
    \begin{bmatrix}
        x_{0,... ,M}& 
        y_{0,... ,M}& 
        v_{0,... ,M}&
        a_{0,... ,M}&
        \delta_{0,... ,M}
    \end{bmatrix}.
\end{equation}
Here $x,y$ being the coordinates of the vehicle's trajectory, $v$ being the vehicle's velocities, $a$ being the vehicle's accelerations and $\delta$ being the vehicle's steering angles. The observation encompasses all $M$ actors present in the scenario, with values recorded from the initial time step until the final time step of the scenario. The state representation $\bm{z}$ is derived by concatenation of the feature vector $f$, achieved goal $g_{a}$ and desired goal $g_{d}$. The GRU is simultaneously trained during the policy update step.

\subsection{Constraint-based Goal Definition}
\label{subsec:algo_goal_space}

In goal-conditioned \ac{RL} goals are defined as desired properties or features. 
A goal is therefore asking the agent to achieve a final states that suffices specific features. 
Training the goal-conditioned \ac{RL} agent requires defining a goal-conditioned reward function $r(\bm{s}, \bm{g})$ that indicates whether the desired state is reached. 
Typically, the completion of a goal is a criterion that is known in advance. 
In the context of safety-critical scenario generation, scenarios must meet specific conditions tailored for targeted training and evaluation of \ac{ADS}. As outlined in \cite{menzel2018scenarios}, scenarios can be categorised according to their level of abstraction, namely functional, logical and concrete. Logical scenarios are defined by parameter ranges, while concrete scenarios are defined by single parameter values. 
Both approaches fail to consider the inter-dependencies and relationships between the attributes and behaviours of the actor occurring in the scenario.
Given that GOOSE operates at the scenario level, with the objective of exerting control over the behaviour of actors through the curse of the scenario, we are particularly interested in the relationships rather than modifying parameters at the initial state. 
\cite{neurohr2021criticality} address the relationships within traffic, particularly cause-and-effect relationships, by defining abstract scenarios expressed through the use of constraints.
The OpenScenario DSL \cite{ASAM_OpenSCENARIO_DSL} defines abstract scenarios through the use of equality and inequality constraints.

The combination of the goal definition of goal-condi- tioned \ac{RL} and the requirements of safety-critical scenario generation allows for the definition of goals as desired abstract scenario properties that fulfill the test objective. While the \ac{RL} agent functions as a constraint solver to explore the space of unknown scenarios. 
%and have physically plausible vehicle behaviour.
In the context of goal definition, two distinct types of goals can be identified: equality goals, denoted by $\bm{g}_{eq}$, and inequality goals, denoted by $\bm{g}_{ieq}$. 
Equality goals are concerned with the explicit achievement of specific values, such as certain vehicle distances or positions. 
In contrast, inequality goals focus on complex relations within traffic, such as acceleration, steering angle or criticality measures.   
Accordingly, the following definition of goals is proposed:
\begin{equation}
    \bm{g} = 
    \big[
    \underbrace{c_1,...,c_n}_{\bm{g}_{eq}},
    \underbrace{d_1,...,d_m}_{\bm{g}_{ieq}}
    \big].
\end{equation}
\begin{algorithm}[!b]
    \caption{Training process of GOOSE}\label{alg:rl_training}
    \begin{algorithmic}[1]
        \STATE \textbf{Initialize:} Scenario simulation, replay buffer $\mathcal{D}$, goal-conditioned policy $\pi_{\theta}(\cdot|s,g)$%, Q-functions $Q_{\psi}(a,s,g)$, 
        \FOR{each episode}
            \STATE Reset environment and scenario simulation
            \STATE Sample initial state $\bm{s}_0 \sim \rho_0$, desired goal $\bm{g} \sim \rho_g$
            \STATE $t \gets 1$
            \WHILE{$t \leq T$}
                \STATE Sample action $\bm{a}_t \sim \pi_{\theta}(\bm{s}_t, \bm{g})$ 
                \STATE Apply $\bm{a}_t$ to modify the scenario (Sec. \ref{subsec:algo_action_space})
                \STATE Step environment by running the scenario simulation $\bm{s}_{t+1} \sim p(\bm{s}_{t+1} \vert \bm{s}_t, \bm{a}_t)$
                \STATE Compute reward $r_t$ for policy $\pi_{\theta}$ based on the goal definition (Sec. \ref{subsec:algo_goal_space}) 
                \STATE Store $(\bm{s}_t, \bm{a}_t, r_t, \bm{s}_{t+1}, \bm{g})$ in $\mathcal{D}$
                %\STATE Update Q-networks $Q_{\theta}$
                \IF{$t \geq$ $T$ \OR goal reached}
                \STATE End episode
                \ENDIF
                \STATE $t \gets t + 1$
            \ENDWHILE
            \STATE Update policy network $\pi_{\theta}$ to maximize $r_t$ based on preprocessed state $\bm{z}_t$ (Sec. \ref{subsec:algo_observation_space})
        \ENDFOR
        \STATE \textbf{output} agent policy $\pi_{\theta}$
    \end{algorithmic}
\end{algorithm}
Here $c_{1,...,n}$ are the $n$ equality constraints and $d_{1,...,m}$ are the $m$ inequality constraints that are required to be satisfied. 
The proposed goal representation is utilized to derive a reward function that indicates whether the desired scenario properties have been fulfilled.
The Euclidean distance between the state and the equality goal is used as the first metric, while the fulfillment of the inequality goal is used as the second metric.
In summary, the following reward function results:
\begin{equation}
    r(\bm{s},\bm{g}) = -  \delta \big(\| \phi_{eq}(\bm{s}) - \bm{g}_{eq}\| \geq \epsilon \big)
     \times \delta \big(\phi_{ieq}(\bm{s}) \geq \bm{g}_{ieq} \big)
\end{equation}
where $\delta$ is the indicator function, $\epsilon \in \mathbb{R}^{+}$ is the distance threshold and $\phi_{eq}, \phi_{ieq}$ are mapping functions that map the observation to goals. In general, agents receive a negative reward of $-1$ at all time steps until $\phi_{ieq}(\bm{s}) < \bm{g}_{ieq}$ and $\|\phi_{eq}(\bm{s}) - \bm{g}_{eq}\| < \epsilon$, after which a reward of $0$ is provided and the episode terminates. In order to minimize the overall negative reward, this approach encourages the goal to be achieved as quickly as possible.

\subsection{Algorithm Summary}
\label{subsec:algo_summary}

The entire pipeline of \ac{GOOSE} is summarized in Algorithm \ref{alg:rl_training}. 
We train the goal-conditioned policy on randomly selected initial states and goals sampled from the distribution of desired goals $\rho_g$. At test time a new initial state is presented and the trained policy is employed to direct the scenario towards the desired scenario properties.

\section{Experiments and Results}

As a test bed, we use a set of three abstract scenarios based on the UN Regulation No. 157 for Active Lane Keeping Systems (ALKS) \cite{ALKS_2022}. 
We define \ac{RL} policies to control the trajectory of a selected vehicle, while the ego vehicle aims to maintain a safe state, and all other vehicles follow their predetermined trajectories. 
The action space of the agent is continuous and specifies the desired change of the control points and weights of the \ac{NURBS} curve. 
In every environment, a cubic \ac{NURBS} curve ($p=3$) with five control points ($n=5$) is utilized. 
\begin{figure}[!b]
    \centering
    \includegraphics[width=0.99\linewidth]{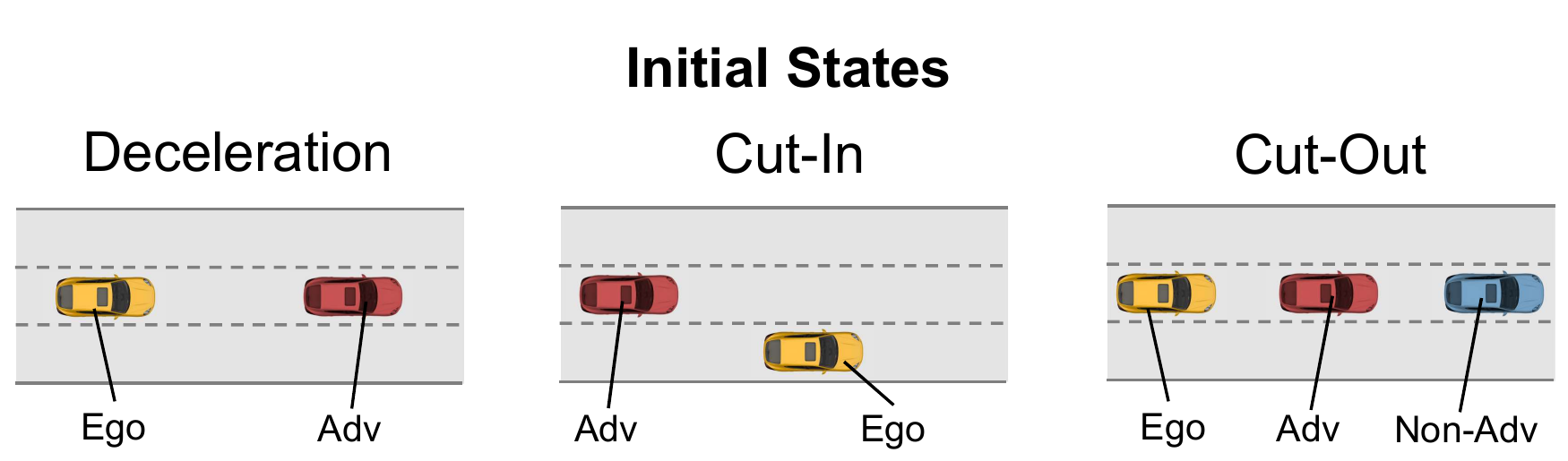}
    \caption{Initial states for the deceleration, cut-in and cut-out scenarios}
    \label{fig:initial_states}
\end{figure}
In the first environment, the \ac{RL}-controlled vehicle should decelerate in front of the ego vehicle, which is therefore designated as \texttt{deceleration}. 
The adversarial vehicle and the ego vehicle are initiated in the same lane, with a longitudinal distance between 20 and 40 meters between them (Fig. \ref{fig:initial_states}).  
The goal $g$ includes one equality constraint $c_1$ and two inequality constraints $d_1, d_2$. 
Constraint $c_1$ is concerned with the distance between the ego and the adversarial vehicle, whereas the $d_1$ and $d_2$ are related to the acceleration and steering of the adversarial vehicle. 
The second environment is called \texttt{cut-in}, wherein the \ac{RL}-controlled vehicle should merge in front of the ego vehicle. 
The scenario is initiated by positioning the adversarial vehicle to the left or right of the ego vehicle, the side from which the cut-in is executed. 
The longitudinal distance between the ego vehicle and the adversarial vehicle is set between 0 and 60 meters (Fig. \ref{fig:initial_states}). 
The goal $g$ is analogous to the deceleration environment, but a third inequality constraint $d_3$ has been incorporated to address the angle between the ego and the adversarial vehicle. 
This is done to prevent the adversarial vehicle from crashing into the ego.  
The last environment is called \texttt{cut-out}. The objective is that the \ac{RL}-controlled vehicle should exit the lane of the ego vehicle. 
All vehicles are initiated in the same lane, with a longitudinal distance between 20 and 40 meters between adversarial vehicle and the ego vehicle and a longitudinal distance of between 20 and 60 meters between adversarial vehicle and the non-adversarial vehicle (Fig. \ref{fig:initial_states}). 
The goal $g$ includes one equality constraint $c_1$ and three inequality constraints. 
Constraint $c_1$ is concerned with the distance between the ego and the non-adversarial vehicle. 
The constraints $d_1$ and $d_2$ are related to the acceleration and steering of the adversarial vehicle. Finally, $d_3$ addresses the distance between adversarial, non-adversarial and ego vehicle.
All environments have a episode length of $T_{max}=200$ steps and will terminate if the agent reaches the desired scenario properties. 
It is considered a success, if the agent reaches within $\epsilon$ of the desired goal. 
All tasks are associated with sparse rewards, which are defined as either 0 or -1.

\subsection{Implementation Details}

\subsubsection{\ac{RL} Algorithm} 
To solve the goal-conditioned scenario generation task, GOOSE utilizes Dropout Q-Functions (DroQ) \cite{hiraoka2021dropout, smith2022walk}, an off-policy model-free reinforcement learning algorithm that enhances \ac{SAC} \cite{haarnoja2018soft} through the integration of dropout and layer normalization into the Q-function networks. \ac{SAC} is a model-free \ac{RL} method that uses entropy to increase the diversity of action.
\ac{HER} \cite{andrychowicz2017hindsight} is used to improve the sample efficiency of the goal-conditioned policy. The hyperparameters are listed in Table \ref{tab:rl_parameters}.

\begin{table}[!b]
    \centering
    \caption{DroQ hyperparameters}
    \label{tab:rl_parameters}
    \begin{tabular}{l|c}
        \hline
        & \\[\dimexpr-\normalbaselineskip+1pt]
        Hyperparameter & Value \\
        & \\[\dimexpr-\normalbaselineskip+1pt]
        \hline
        & \\[\dimexpr-\normalbaselineskip+1pt]
        optimizer & \texttt{Adam} \\
        number of samples per batch & \num{256} \\
        learning rate ($\lambda$) & \num{3e-4} \\
        gradient steps & \num{4} \\
        policy delay & \num{2} \\
        replay buffer size & \num{1e6} \\
        replay buffer class & \texttt{HER} \\
        goal selection strategy & future \\
        number of sampled goals & 4 \\
        discount ($\gamma$) & \num{0.95} \\
        entropy target & $-\dim(\bm{a})$\\
        entropy temperature factor ($\alpha$) & \num{1} \\
        target network smoothing coefficient ($\rho$) & \num{5e-3} \\
        number of hidden layers & \num{2} \\
        number of hidden units per layer & \num{256} \\
        number of GRU layers & \num{1} \\
        number of hidden units per GRU layer & \num{128} \\
        activation functions & \texttt{ReLU} \\
        dropout rates & \num{0.02} \\
        \hline
    \end{tabular}
\end{table}

\subsubsection{Simulator and Dataset} For the generation and alteration of traffic scenarios, the ScenarioGym framework is selected \cite{scott2023scenario}. As all our scenario generation environments are map-agnostic, a multi-lane scenario from the Argoverse 2 motion forecasting dataset \cite{wilson2023argoverse} was utilized for all experiments. To maintain an observation vector of constant length during scenario generation, the simulation time is set to 10 seconds with a time step of 0.1 seconds. A starting position that meets the starting conditions is randomly selected on the map. For example, if the cut-in side is specified as left, there must be a left lane next to the ego vehicle.

\subsubsection{System under Test} The intelligent driver model (IDM) \cite{treiber2000congested} is applied to simulate the longitudinal dynamics, including the acceleration and deceleration of the ego vehicle. The IDM parameters are listed in Table \ref{tab:idm_parameters}.

\begin{table}[!t]
    \vspace{3mm}
    \centering
    \caption{IDM parameters \cite{treiber2000congested}}
    \label{tab:idm_parameters}
    \begin{tabular}{l|c}
        \hline
        & \\[\dimexpr-\normalbaselineskip+1pt]
        Parameter & Value \\
        & \\[\dimexpr-\normalbaselineskip+1pt]
        \hline
        & \\[\dimexpr-\normalbaselineskip+1pt]
        desired velocity ($v_{0}$) & \num{15.0} \si{\metre\per\second}\\
        safe time headway ($t_{HW}$)& \num{1.6} \si{\second}\\
        maximum acceleration ($a$) & \num{0.73} \si{\metre\per\square\second}\\
        comfortable deceleration ($b$) & \num{1.67} \si{\metre\per\square\second}\\
        acceleration exponent ($\delta$) & \num{4} \\
        minimum distance ($s_{0}$)  & \num{2.0} \si{\metre}\\
        \hline
    \end{tabular}
\end{table}

\subsection{Training Performance}

\begin{figure}[!b]
\centering
\vspace{-1mm}
  \begin{minipage}{0.49\columnwidth}
    \centering
    \includegraphics[width=\linewidth]{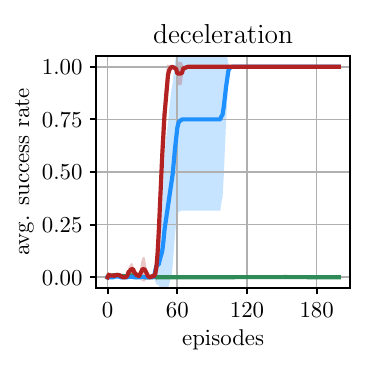}
  \end{minipage}
  \begin{minipage}{0.49\columnwidth}
    \centering
    \includegraphics[width=\linewidth]{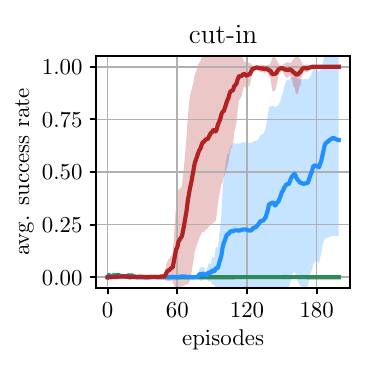}
  \end{minipage}\par
  \vspace{-2.5mm}
  \begin{minipage}{0.49\columnwidth}
    \centering
    \includegraphics[width=\linewidth]{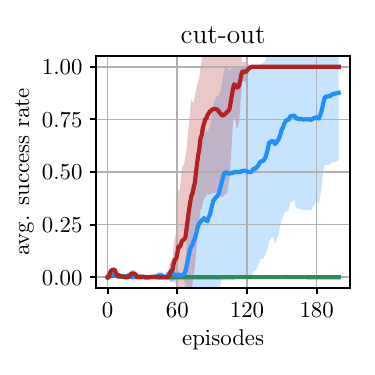}
  \end{minipage}
  \begin{minipage}{0.49\columnwidth}
    \centering
    \includegraphics[width=\linewidth]{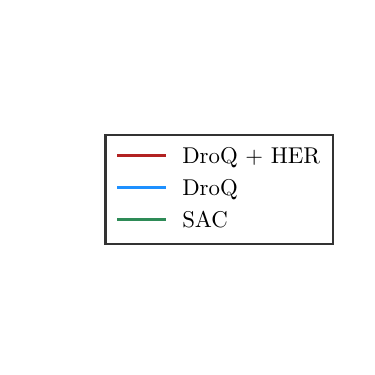}
  \end{minipage}
\caption{Training success rates of the agent on the 3 generation environments \texttt{deceleration}, \texttt{cut-in} and \texttt{cut-out}. Curves are averaged over multiple seeds and shaded regions represent one standard deviation.}
\label{fig:success_rates}
\end{figure}

The goal-conditioned agent is trained according to Algo. \ref{alg:rl_training}. \ac{GOOSE} uses DroQ and \ac{HER} to optimize over $4$ subgoals.
To demonstrate the overall performance we compare the training success rates on the three scenario generation environments to DroQ and \ac{SAC}. 
The results are shown in Fig. \ref{fig:success_rates}. 
\ac{GOOSE} outperforms baselines in all tasks. 
Our method quickly learns to solve the scenario generation tasks at the beginning of the training, demonstrating high data efficiency and low variance across different training seeds. \ac{SAC} fails in all three environments since they have very sparse rewards. DroQ is capable of achieving success in some seeds, although the average success is characterised by high variance and slower convergence.  

\subsection{Goal-conditioned Scenario Generation}

\begin{figure*}[!t]
\centering
\vspace{3mm}
  \begin{minipage}{0.333\textwidth}
    \centering
    \includegraphics[width=0.935\linewidth]{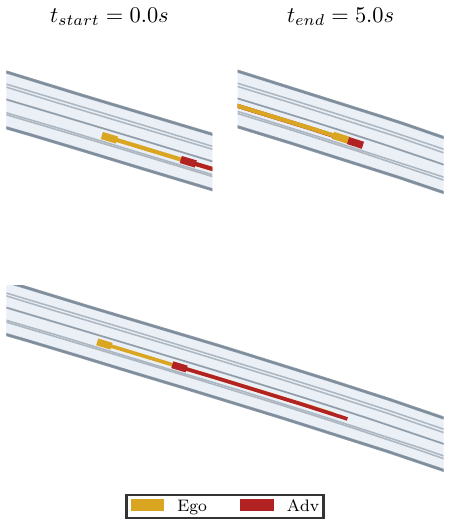}
    \footnotesize (a) Deceleration
  \end{minipage}%
  \begin{minipage}{0.333\textwidth}
    \centering
    \includegraphics[width=0.935\linewidth]{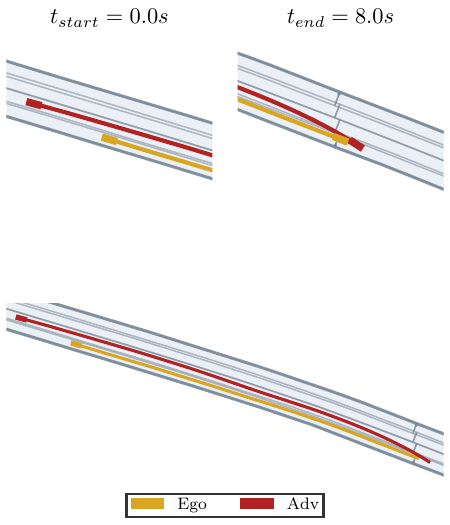}
    \footnotesize (b) Cut-in
  \end{minipage}%
  \begin{minipage}{0.333\textwidth}
    \centering
    \includegraphics[width=0.935\linewidth]{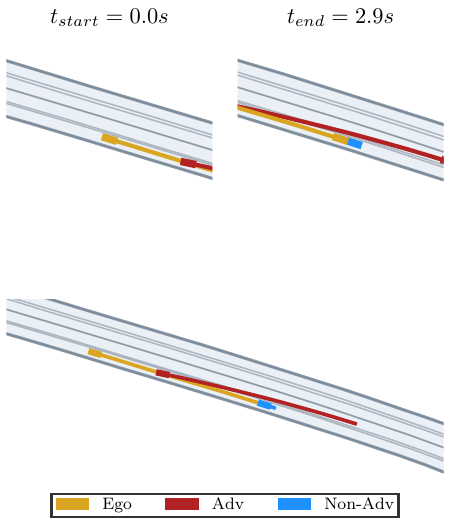}
    \footnotesize (c) Cut-out
  \end{minipage}
\vspace{-1mm}
\caption{Qualitative generation results of the goal-conditioned policy in the \texttt{deceleration}, \texttt{cut-in} and \texttt{cut-out} environments. The scene is visualised by the lane boundary lines as well as the actor trajectories, with the ego vehicle in yellow, the RL-controlled vehicle in red and all other actors in blue. At the top start and final time steps are visualised. The scenarios end when a collision with the ego vehicle occurs.}
\label{fig:generated_scenarios}
\end{figure*}

In goal-conditioned scenario generation, the \ac{RL} agent should converge to a policy able to reach the desired scenario properties. 
The objective is to evaluate the ability of \ac{GOOSE} to generate safety-critical scenarios in the three proposed environments, namely \texttt{deceleration}, \texttt{cut-in} and \texttt{cut-out}.
The results demonstrate that the approach is capable of generating controllable and realistic safety-critical scenarios, which are essential for comprehensive \ac{ADAS} and \ac{ADS} testing. The following goal specifications are employed in the evaluation process. The objective in the \texttt{deceleration} task is defined as  
\begin{equation} \label{eq:decel_goal}
    g
    = \big[
    \begin{matrix}
        0.0\,\text{m} & 8.0\,\frac{\text{m}}{\text{s}^2} & 0.7\,\text{rad}
    \end{matrix}    
    \big],
\end{equation}
where the first value specifies the desired distance between the ego vehicle and the adversarial vehicle, while the second and third values specify the maximum acceleration or deceleration and maximum steering angle of the adversarial vehicle. In the \texttt{cut-in} task, the goal definition (\ref{eq:decel_goal}) is extended by a third inequality constraint:
\begin{equation}
    g
    = \big[
    \begin{matrix}
        0.0\,\text{m} & 8.0\,\frac{\text{m}}{\text{s}^2} & 0.7\,\text{rad} & 0.5\,\text{rad}
    \end{matrix}    
    \big],
\end{equation}
which specifies the maximum angle between the ego vehicle and the adversarial vehicle in the event of a collision, in order to prevent the adversarial vehicle from crashing into the ego vehicle. The objective of the \texttt{cut-out} task is as follows:
\begin{equation}
    g
    = \big[
    \begin{matrix}
        0.0\,\text{m} & 8.0\,\frac{\text{m}}{\text{s}^2} & 0.7\,\text{rad} & 0.25\,\text{m}
    \end{matrix}    
    \big].
\end{equation}
Here the first value specifies the desired distance between the ego vehicle and the non-adversarial vehicle. 
The second and third values indicate the maximum acceleration or deceleration and maximum steering angle of the adversarial vehicle, respectively. 
The fourth value specifies the minimum distance between the adversarial vehicle and all other vehicles. 
At the beginning of each episode, a new random initial state is sampled from the set of potential initial states that fulfill the specified starting conditions (see Fig. \ref{fig:initial_states}). \ac{GOOSE} generates crash scenarios using the outlined goals and the sampled initial state. 

Fig. \ref{fig:generated_scenarios} illustrates the qualitative examples, demonstrating how \ac{GOOSE} can challenge the \ac{SUT} in various driving situations. 
In Fig. \ref{fig:generated_scenarios}a the generated scenario in the \texttt{deceleration} environment is displayed. 
The adversarial vehicle is positioned in front of the ego vehicle at the beginning of the scenario.    
A combination of acceleration and deceleration of the adversarial vehicle results in a collision at $t=5.0$ seconds.
The results of the \texttt{cut-in} environment are presented in Fig. \ref{fig:generated_scenarios}b. 
The adversarial vehicle is positioned behind the ego vehicle in the left lane.  
The adversarial vehicle overtakes the ego vehicle from the left and cuts in the lane of the ego while decelerating, resulting in a collision at $t=8.0$ seconds.
Finally, in Fig. \ref{fig:generated_scenarios}c, the generated scenario in the \texttt{cut-out} environment is depicted. 
The ego vehicle, adversarial vehicle and non-adversarial vehicle are all positioned in the same lane. 
During the scenario simulation, the adversarial vehicle performs a close cut-out manoeuvre, thereby exposing the non-adversarial vehicle to the ego, resulting in a collision between the ego and the non-adversarial vehicle at $t=2.9$ seconds.

\section{Conclusion and Future Work}

We presented \acf{GOOSE}, a goal-conditioned reinforcement learning method that generates safety-critical scenarios by exploring the space of adversarial trajectories to challenge automated driving systems (ADS). 
\ac{GOOSE} uses \acf{NURBS} as a modeling tool for diverse trajectories, while retaining a relatively small set of optimizable parameters. 
This results in a significant reduction of the action space for the goal-conditioned policy. 
\ac{GOOSE} employs a constraint-based goal definition to specify the scenario generation task, which is based on the OpenScenario DSL standard.
Experimental results on scenarios from the UN Regulation No. 157 for \acf{ALKS} show that \ac{GOOSE} successfully generates scenarios that lead to critical events and outperforms other model-free \ac{RL} methods in terms of data efficiency. 
While our approach only controls a single target vehicle, future work could explore multi-agent control for safety-critical scenario generation. 
Another exciting direction for future work is to examine the impact of different criticality measures on the characteristics of the generated scenarios. 
Future work could also apply the goal-conditioned policy to scenarios built from real-world data to increase criticality.

\bibliographystyle{IEEEtran}
\bibliography{IEEEabrv, indices/lit.bib}

\end{document}